\documentstyle[multicol,aps,prb,graphicx]{revtex}
\begin{document}
\newcommand{\achtung}{\marginpar{!}}
\newcommand{\attent}{\marginpar{\rule[0mm]{0.5mm}{8mm}}}
\newcommand{\vnimanie}{\marginpar{\rule[0mm]{0.5mm}{32mm}}}

\title{ Kondo Effect in a Metal with Correlated Conduction Electrons:\\
Diagrammatic Approach}

\author{ M. Neef\( ^{1}
    \),
S. Tornow\( ^{1,*}
    \),
V. Zevin\( ^{2}
     \),
and G. Zwicknagl\( ^{1}
    \)}

\address{
$^{1}$ Institut f\"ur Mathematische Physik, Technische Universit\"at 
Braunschweig, Germany \\
\(^{2}\)The Racah Institute of Physics, The Hebrew
University of Jerusalem, 91904 Jerusalem, Israel}
\date{\today}
\maketitle

\begin{abstract}
We study the low-temperature behavior of a magnetic impurity which is weakly
coupled to correlated conduction electrons. To account for conduction electron 
interactions a diagrammatic approach in the frame of the 1/N expansion
is developed.  The method allows us to study various consequences
of the conduction electron correlations for the ground state and the low-energy
excitations. We analyse the characteristic energy scale in the limit of weak
conduction electron interactions. Results are reported for static properties
(impurity valence, charge susceptibility, magnetic susceptibility, and
specific heat) in the low-temperature limit.
\end{abstract}

\begin{multicols}{2}
\narrowtext
\section{Introduction}

\label{sec:intro}

Metals with strongly correlated electrons exhibit highly complex phase diagrams
at low temperatures reflecting a rich variety of possible ground states. Prominent
examples are the well-known normal Fermi liquid state as well as magnetically
ordered, superconducting and insulating phases which may coexist or compete
within the same material. A key to a quantitative understanding of the unusual
phases is therefore a quantitative description of electronic correlations and
their observable consequences.

The present paper focusses on (dilute) magnetic alloys with correlated conduction
electrons, i.e., we consider host metals with correlated conduction
electrons containing a small amount of magnetic ions. We investigate the question
how conduction electron correlations affect the formation of a nonmagnetic Fermi
liquid ground state commonly referred to as Kondo effect. The latter has been
known to be the source of many anomalous properties in magnetic alloys with
noninteracting conduction electrons. In addition to its relevance in magnetic
alloys the Kondo effect is becoming important in the study of interacting mesoscopic
systems. Theoretical techniques which provide a detailed quantitave understanding
of the physical properties of these systems are hence highly desirable. To leading
order in the low impurity concentration the electronic properties of dilute
magnetic alloys can be calculated in two steps. First one has to determine the
elctronic properties of the host which will not be significantly affected by
the addition of a small amount of impurities. In the second step the contribution
of the magnetic ions has to be calculated.

For a metal with uncorrelated conduction electrons the first part of the problem
is solved by standard methods of electronic structure calculation. The theory
for the second step is well established 
\cite{Tsvelick83,Andrei83,Fulde88a,Zwicknagl92,Hewsonbook}.
The theoretical techniques available include exact solutions for equilibrium
properties as well as approximate methods for dynamic properties. Of particular
importance in this context is the diagrammatic approach based upon the large-degeneracy
expansion. This scheme can be generalized to the treat non-equilibrium properties
which makes it a very flexible method.

The central goal of the present paper to extend the large-degeneracy expansion
for the normal-state properties of dilute magnetic alloys to the case of host
metals with correlated conduction electrons. In this case, the first step, i.
e., the treatment of the host is a highly non-trivial problem which has not
yet been solved. Partial answers, however, can be found in limiting
cases. 
Although the diagrammatic approach developped in the present paper is valid
for a general conduction electron interaction ( CEI ), we
focus  on systems  where the ground state and the low-energy excitations
of the interacting conduction electrons smoothly evolve from those of the non-interacting
reference system. This is in marked contrast to the specific behavior encountered
in one-dimensional systems. Theoretical studies which have been performed for
various models including both impurity spins 
\cite{Lee92,Furusaki94,Schiller95}
and Anderson impurities \cite{Li95,Phillips96,Schiller97} coupled to Luttinger
liquids predict rich phase diagrams. Adopting well-established models for the
electronic properties of the host, we calculate the evolution of the characteristic
energy \( T_K \) of the low-lying magnetic excitations with the conduction
electron repulsion. In the case of uncorrelated conduction electrons the latter
is usually much smaller than the typical energy scale of the conduction electrons
set by the band width \( D \) and depends exponentially on the inverse coupling
between the localized electron and the extended conduction states. This fact
is a direct consequence of the Fermi liquid ground state realized in normal
metals. The diagrammatic approach allows us to explicitly and quantitatively
study how the different consequences of electronic correlations (mass renormalization,
effective interactions etc) affect the Kondo effect.

The main scope of this paper is to analyze how CEI influence 
the contribution of magnetic impurities to
measurable properties 
 in general and its scaling 
properties in particular.
We calculate thermodynamic properties (impurity valence, charge
susceptibiliy, magnetic susceptibility and specific heat) in the
low-teperature limit to leading order in the inverse degeneracy. 

Recent calculations for a magnetic impurity in a metal with
interacting conduction
electrons \cite{Davidovich98,Hofstetter00} adopted either the DMFT approach
\cite{Georges96a} or the NRG but for a very special model \cite{Takayama98}. 
The model calculations mentioned above predict non-trivial
variation with the Coulomb repulsion of the characteristic temperature 
\( T_K \).

Generally, the modifications introduced by the conduction electron 
interactions (CEI ) into the low-energy excitations arise from the
subtle interplay of three
different types of influences. First, the density of conduction states at the
Fermi level is changed. Second, the probability for virtual
transitions between
impurity and conduction states are reduced by the on-site Coulomb interaction
\( U \). Third, the effective spin coupling between the conduction and 
impurity electrons is enhanced by the increased number of 
uncompensated spins in the
correlated conduction electron system. Considering these facts, it is
not surprising
that model studies accounting only for selected aspects arrive at
rather controversial
conclusions concerning the Kondo effect in metals with correlated electrons
\cite{Khaliullin95,Schork96,Itai96}. The Kondo spin model generalised
to the case of the interacting conduction-electron host was discussed
in \cite{Khaliullin95}
and it was shown there that two-particle Green's functions of host electrons
( vertex corrections ) are an essential component of the theory which leads
to an enhancement of the \textit{exponential} Kondo scale for a \textit{weak}
CEI. This enhancement may be traced to the third type of effects caused by the
CEI. The ground state energy of the Anderson impurity for \textit{weak} CEI
was considered in the frame of \( 1/N \) expansion \cite{Schork96}. The same
enhancement of the exponential Kondo scale, formally due to the 
renormalization increase of the hybridization width \( \Delta  \), appears 
in this work \cite{Schork96}.
In contrast to the above-mentioned findings a decrease of \( T_K \) due to
the CEI in the Hubbard model was reported in the paper \cite{Itai96}.
This decrease is a consequence of the change in the single electron properties
of conduction electrons caused by the interaction \( U \) ( including the change
of the chemical potential as the function of \( U \) ~). The vertex corrections
influence which renormalizes both parameters of the Anderson impurity model\cite{Tornow97},
\( \epsilon _f \) and \( \Delta \) are not considered in \cite{Itai96}.
At this point, we should like to mention that the role of the Coulomb interaction
between the magnetic impurity electron and conduction electrons, \( U_{fc} \),
was broadly discussed. We do not discuss here the Coulomb interaction between localized and
conduction electrons which is considered in its various aspects in 
\cite{Costi91,Takayama93,Perakis93,Giamarchi93,Bauer95}. It was shown that its effect
at \( U_{fc}\ll U_f \) may be fully absorbed by the renormalization of two
parameters of the Anderson impurity Hamiltonian: the impurity electron energy
level \( \epsilon _f \) and the hybridization width \( \Delta  \). In the following
we assume that the on-site impurity electron Coulomb repulsion \( U_f \)
is very large, \( U_f\rightarrow \infty  \), and we do not take into account
explicitly the \( U_{fc} \) interaction.

The paper is organized as follows: In Section II we begin with a discussion   
of the Hamiltonian for an Anderson impurity embedded in a metallic host with
correlated conduction electrons and the extension of the standard selfconsistent
large-degeneracy approximation to the case of CEI. Both the interaction-induced
changes in the single-electron spectral function of interacting
conduction electrons
and their vertex function are included. In Section III expressions for configurational 
selfenergies together with the NCA integral equations are formulated for a general 
case of CEI. The expressions are evaluated
in Section IV for a model where the Coulomb vertex function is only weakly 
frequency-dependent. Thermodinamic properties at zero temperature are presented in Section V
and  Section VI contains discussions
and summary. Technical details related to the explicit evaluation of diagrams and discussions
of the fourth-order hybridisation coupling are 
in the appendices. Some of the results appeared in the
short unpublished preprint \cite{Tornow97}.

\section{Model and calculational scheme}

\label{sec:model} We adopt a generalized Anderson model for a magnetic 
impurity coupled to interacting conduction electrons. The resulting 
Hamiltonian reads
\begin{equation}
H=H_{band}+H_{imp}+H_{mix}
\end{equation}
where the three components describe the conduction electrons, the f
states and a hybridization or mixing interaction between the two
\begin{eqnarray}
H_{band} & = & \sum _{\vec{k}\sigma }\epsilon _{\vec{k}}
c_{\vec{k}\sigma }^{\dagger }c_{\vec{k}\sigma }+H_{CEI}\nonumber \\
H_{imp} & = & \sum _{m}\epsilon _fn_{fm}+{U_f\over2 }
\sum _{m\neq m{\prime }}n_{fm}n_{fm^{\prime }}\nonumber \\
H_{mix} & = & \sum _{\vec{k},m,\sigma }\left( V_{m\sigma }(\vec{k})
f_{m}^{\dagger }c_{\vec{k}\sigma }+h.c.\right) \quad .\label{eq:H} 
\end{eqnarray}
The creation (annihilation) operators for conduction electrons with momentum
\( \vec{k} \), band energy \( \epsilon _{\vec{k}} \) and spin \( \sigma  \)
are denoted by \( c_{\vec{k}\sigma }^{\dagger }(c_{\vec{k}\sigma })
\). Throughout this paper, all energies are measured relative to the 
Fermi level. The conduction states are assumed to be orbitally 
non-degenerate. Their interaction is accounted for by 
\begin{eqnarray}
H_{CEI}&=&{1\over 2L}\sum _{\vec{k},\vec{k}^{\prime },\vec{q}\ \sigma,
\sigma ^{\prime }}
 U(\vec{k}+\vec{q},\vec{k}^{\prime }-\vec{q};\vec{k}^{\prime },\vec{k})
\nonumber \\
&& \;\;\;\;\;\;\;\;\;\;\;\;\;\;\;\;\;\;\;
\times c_{{\vec{k}+\vec{q}}\sigma }^{\dagger }
c_{\vec{k}^{\prime }-\vec{q}\sigma ^{\prime }}^{\dagger }
c_{\vec{k}^{\prime }\sigma ^{\prime }}c_{\vec{k}\sigma }.
\label{eq:H_{C}EI}
\end{eqnarray}
 where \( L \) is the number of lattice sites. In the present paper, we
approximate \( H_{CEI} \) by a Hubbard-type interaction 
\begin{equation}
U(\vec{k}+\vec{q},\vec{k}^{\prime }-\vec{q};\vec{k}^{\prime },\vec{k})
\rightarrow U
\end{equation}
where \( U \) denotes the local Coulomb repulsion between two
conductionelectrons at the same lattice site. Another important
example which shall be studied in a forthcoming paper are Fermi liquid systems
where the CEI renormalizes the quasiparticle dispersion 
\( \epsilon _{\vec{k}} \) and also introduces a `residual' interaction 
among them.

The \( f_{m}^{\dagger }(f_{m}) \) are the creation (annihilation) operators
for \( f \)-electrons at the impurity site. They are characterized by 
the total angular momentum \( J \) and a quantum number \( m \) which 
denotes the different states \( m=1,\dots ,N \) within the 
\( N \)-fold degenerate ground state
multiplet with orbital energy \( \epsilon _f \). The Coulomb repulsion 
\( U_f \) between two
\( f \)-electrons at the impurity site is assumed to be much larger than the
other energy scales and therefore we may let \( U_f\rightarrow \infty  \).
For simplicity we do not include here excited multiplet states, ignore crystal
electric field splittings and assume that the impurity has only one electron
( hole ~) in its magnetic configuration. We account for the large
Coulomb interaction
among the f-electrons \( U_f\rightarrow \infty  \) by restricting the Hilbert
space, i. e., by removing all states in which the f occupancy exceeds
unity.

The mixing between the two subsystems is conveniently characterized by the
''hybridization width''\cite{CommentHybridizationWidth}
\begin{equation}
\Delta(E)= \pi \frac{1}{L} \frac{1}{N}\sum_{\vec{k}\ \sigma \ m}
\left|V_{m\sigma}(\vec{k}) \right|^2 \ \delta(E-\epsilon_{\vec{k}})
\quad .
\end{equation}
We are mainly interested in the regime 
 $ \left| \epsilon_f \right| \gg \Delta_m(0) $
which is usually referred as ''local moment regime''.

\begin{figure}[bt]
\begin{center}
\includegraphics[width=8.0cm]{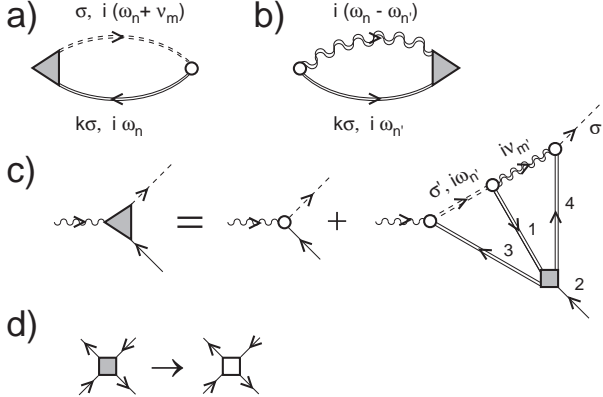}
\end{center}
\vspace{10pt}
\caption{Self-consistent f configuration self-energies and
contributions to the vertex. The solid, dashed and wavy lines
represent the dressed propagators for conduction electrons, occupied
and empty f states. The open circle denotes the bare hybridization V
while open and filled squares are the bare on-site Coulomb repulsion
and the two-particle vertex 
$\Gamma_{\sigma,\bar{\sigma}}^{(U)}(1,2;3,4 )$, respectively. 
(a) Empty state self-energy \(\Sigma_0(i\nu_{m})\). 
(b) Occupied state self-energy \(\Sigma_{\sigma}(i\omega_n)\). 
(c) Contribution to the effective hybridization vertex.
(d) Lowest order correction. }
\label{fig:BubbleSelf}
\end{figure}

The central goal is to calculate the impurity contribution to the
low-energy properties of the dilute magnetic alloy. The latter are
given in terms of the Green's functions for the empty f-state 
\( |0>\) (\( 4f^0 \)- configuration)
and the occupied \( f \) states \( |m> \) 
(\( 4f^{1} \)- configuration) denoted
by \( G_0(z) \) and \( G_{m}(z) \), respectively, 
\begin{equation}
\label{eq:Green}
G_0(z)\; =\; {1\over {z-\Sigma _0(z)}}\qquad ;\qquad G_{m}(z)\;
=\; 
{1\over {z-\epsilon _f-\Sigma _{m}(z)}}\quad .
\end{equation}
They are coupled through the configurational selfenergies 
$\Sigma _0(z)$ and  and $\Sigma _m(z)$ for which we derive
expressions proceeding in close
analogy to the case of non-interacting conduction electrons.

The electronic properties of the metallic host are 
not affected by the presence of a small amount of magnetic impurities.
To leading order in the small concentration they are characterized by
the $1$- and $2$-particle Green's function describing the single-particle
excitations and the two-particle correlations of the interacting
conduction electrons, respectively.

We assume that the 
single-electron Green's function 
\begin{equation}
\label{eq:GSigmac}
G_{\sigma }(\vec{k},z)=\frac{1}{z-\epsilon _{\vec{k}}-
\Sigma _{\sigma }(\vec{k},z)}
\end{equation}
 as well as the conduction electron selfenergy 
\( \Sigma _{\sigma }(\vec{k},z) \)
do not explicitly depend upon the 
wave vector \( \vec{k} \) but vary with \( \vec{k} \) mainly through 
the bare band energy, i.e. 
\begin{equation}
\label{eq:Gkdependence}
G_{\sigma }(\vec{k},z)=G_{\sigma }(\epsilon _{\vec{k}},z)\qquad
;\qquad 
\Sigma _{\sigma }(\vec{k},z)=\Sigma _{\sigma }(\epsilon _{\vec{k}},z)\quad .
\end{equation}
This condition is always satisfied in the DMFT approach \cite{Georges96a}
where the dominant many-body effects are included in a local selfenergy. As
a consequence, also the general n-particle Green's functions of the conduction
electron system depend upon the wave vectors through the corresponding band
energies.

The configurational self-energies \( \Sigma _0 \) and 
\( \Sigma _{m} \) are derived by means of a perturbation expansion 
in terms of Green's functions for the 
f-configurations, 1- and 2-particle
Green's functions for the interacting conduction electrons as well as (bare)
hybridization vertices. The rules for constructing and evaluating 
the empty- and occupied-state selfenergies
in the restricted Hilbert space of the infinite-U-Anderson model are 
summarized in \cite{Bickers87b}. Typical contributions to the 
f-configurational selfenergies
are displayed in Figure \ref{fig:BubbleSelf}. These include the non-crossing
diagrams Figure \ref{fig:BubbleSelf}(a) und (b) where the conduction electron
interactions enter through the fully renormalized conduction electron 
propagator. The diagrams \ref{fig:BubbleSelf}(c) and (d) describe
vertex corrections.
We shall show below that under the assumptions 
Eq. (\ref{eq:Gkdependence}) the infinite-order summation of these
diagrams based on the self-consistent approximation for the 
empty- and occupied-f-state propagators can be considered as
the leading order contribution in the inverse degeneracy $1/N$.
The remaining part of the present section is devoted to the justification of
this conjecture.

We start by briefly summarizing the basic facts on which the 
large-degeneracy expansion is based in the
case of non-interacting conduction electrons. The classification
scheme exploits the fact that the bare conduction 
electron propagator 
$1/(z-\epsilon_{\vec{k}})$ 
depends upon the wave vector $\vec{k}$ through the bare band energy. 
As a consequence, the summations over internal $\vec{k}$-vectors
can be decomposed into integrals over
the (bare) band energies \( \epsilon  \) and averages over constant
energy surfaces according to 
\begin{equation}
\label{eq:ksum}
{1\over L}\sum _{\vec{k}}\dots \rightarrow \int d\epsilon \, \, 
{1\over L}\sum _{\vec{k}}\delta (\epsilon -\epsilon _{\vec{k}})\dots =
\int d\epsilon \, \, N(\epsilon )\langle \dots \rangle _{\epsilon }
\end{equation}
Here \( N(\epsilon ) \)
is the density of bare band energies.
 The \( \vec{k} \)-averages \( \langle \dots \rangle _{\epsilon } \) which
contain only combinations of the hybridization matrix elements 
\begin{eqnarray}
&& <V^{*}_{m\sigma }(\vec{k})V_{m'\sigma }(\vec{k})>_{\epsilon }=
\\
&& {1\over {N(\epsilon )}}
\left\{ {1\over L}\sum _{\vec{k}\sigma }V^{*}_{m\sigma
}(\vec{k})V_{m'\sigma }(\vec{k})
\delta (\epsilon -\epsilon _{\vec{k}})\right\}\sim 
V^{2}_{m}(\epsilon )\delta _{mm'} 
\nonumber
\end{eqnarray}
provide the \( m \)-selection rule \cite{Bringer77}
which simplifies the structure of the selfenergy contributions and 
ultimately allows for
a classification with respect to the small parameter \( 1/N \). 

>From the preceding discussion it is apparent that the assumption 
Eq. (\ref{eq:Gkdependence}) guarantees the validity of the $1/N$ 
classification 
scheme for all diagrams where the conduction electron properties enter via 
the single-particle Green's function. Within this subclass the contributions
displayed in Figure \ref{fig:BubbleSelf}(a) and (b) ( without vertex corrections ) 
are the leading ones 
with respect to the small parameter $1/N$.

To assess the validity of the $1/N$ expansion is more subtle for the 
diagrams containing the two-particle Green's function. Here the simplifying
assumptions Eq. (\ref{eq:Gkdependence}) imply ( see Section III ) that the hybridization 
matrix elements enter 
diagrams in Figure \ref{fig:BubbleSelf}( a,c ) and ( b,c ) in the combination
\begin{eqnarray}
&& V^{(4)}(E_1\sigma_1,E_2\sigma_2;E_3\sigma_3,E_4\sigma_4) = 
\frac{1}{L^{3}}\sum _{\vec{k}_1,\vec{k}_2,\vec{k}_3,\vec{k}_4}\nonumber\\
 &  & \delta (E_1-\epsilon(\vec{k}_1))
\delta (E_2-\epsilon({\vec{k}_2})) \delta (E_3-\epsilon({\vec{k}_3}))
\delta (E_4-\epsilon({\vec{k}_4}))\nonumber\\
 &  & \sum_{m,m'} \ V_{m\sigma_1}(\vec{k}_1)V_{m'\sigma_2}(\vec{k}_2)
V_{m\sigma_3}^{\ast}(\vec{k}_3)
V_{m'\sigma_4}^{\ast }(\vec{k}_4) \nonumber\\
 & & \delta ^{*}(\vec{k}_3+\vec{k}_4-\vec{k}_1-\vec{k}_2)
\quad.
\label{eq:VFour} 
\end{eqnarray}
 where the Laue function 
\begin{equation}
\label{eq:Laue_fnctn}
\delta ^{*}(\vec{k}_1\!+\!\vec{k}_2\!-\!
            \vec{k}_3\!-\!\vec{k}_4)=
\frac{1}{L}\sum _{\vec{R}_n}
\exp \left\{ i\left( \vec{k}_1\!+\!\vec{k}_2\!-\!
            \vec{k}_3\!-\!\vec{k}_4\right) \right\} 
\end{equation}
accounts for momentum conservation up to a reciprocal lattice vector.

In the Appendix \ref{sec:sf} we present a detailed model calculation 
for a rare earth impurity hybridized with tight-binding \( s \)-band states.
The results show that the new contributions to \( \Sigma _0 \)
and \( \Sigma _{m} \) are \( O(1) \) and \( O(1/N) \), i.~e.~, of the same order
of magnitude with respect to $1/N$ as their NCA counterparts. It is 
interesting to note that 
the dominant contributions to $V^{(4)}$ are non-local coming from the 
coupling of the $f$-states to the conduction electrons at the
neighboring sites.

To summarize, the configurational selfenergies displayed
in Figure \ref{fig:BubbleSelf} provide a consistent extension of the well-known
selfconsistent large-degeneracy expansion to the case of interacting conduction
electrons. Actual calculations, however, require the fully renormalized conduction-electron
propagator as well as the Coulomb vertex. Since this problem still remains unsolved
for the Hubbard model \cite{FuldeBook,FazekasBook} we have to adopt 
approximate
expressions derived either from phenomenological considerations or from partial
resummation of selected classes of diagrams.

General qualitative results can be derived in limiting cases. Prominent among
them is the case where the Coulomb vertex can be considered as a static quantity
which includes the limit of weakly interacting conduction electrons as well
as the Fermi liquid case.

\section{Configurational selfenergies}

\label{sec:NCA1}
For non-interacting
conduction electrons,
the self-consistent solution \cite{Coleman83,Bickers87b} has three 
characteristic features: The occupied f-spectrum shifts to peak at a value 
\( {\bar{\epsilon }}_f\simeq \epsilon _f+\Re \Sigma _{m}(\epsilon _f) \),
the dominant contribution to the level shift coming from the continuum
of charge
fluctuations. The resonance in the occupied f-spectral function
acquires a small
width. Finally, the empty state spectral function exhibits a
pronounced structure
at \( \omega _0={\bar{\epsilon }}_f-T_K \) which develops with decreasing
temperature and which sets the scale for the low-temperature behavior. This
feature is the direct manifestation of the Kondo effect reflecting the 
admixture of \( f^0 \)-contributions to the ground state and the 
low-energy excitations.

In this paper, we study the influence of the CEI on this 
non-perturbative feature. Of particular interest are the position 
of the resonance energy \( \omega _0 \) relative to the energy 
\( {\bar{\epsilon }}_f \) of the \( 4f^{1} \) configuration
as well as the weight of the resonance.

Let us first neglect vertex corrections and focus on the 
modifications introduced
by the CEI into the single-particle excitations of the conduction 
electron system.
They are accounted for by inserting the full conduction electron propagator
for interacting electrons \( G_{\sigma }(\vec{k},i\omega _n) \) from Eq.
(\ref{eq:GSigmac}) into the configurational selfenergies 
Figure \ref{fig:BubbleSelf}(a,b).

The selfenergy of the occupied f-level 
\begin{eqnarray}
\Sigma _{m}^{(0)}\left( \omega \right)  & = & \frac{1}{L}\sum
_{\vec{k}\sigma }
V_{m\sigma }(\vec{k})\! \! \int dE\, \, n_f(-E)A_{\sigma}(\vec{k},E)
\nonumber \\
&& \;\;\;\;\;\;\;\;\;\;\;\;\;\;\;\;\;\;\;\;\;\;\;\;
\times G_0\left( \omega -E\right) 
V_{m\sigma }^{*}(\vec{k})\nonumber \\
 & = & \frac{1}{\pi }\! \! \int dE\, \, 
\Delta _{m}^0(E)n_f(-E)G_0(\omega -E)
 \label{eq:fnca}
\end{eqnarray}
is diagonal in $m$ as shown in the previous section.
Here \( n_f(E) \) denotes the Fermi function. The properties of the metallic
host are reflected in the energy-dependent hybridization strength 
\begin{equation}
\label{eq:Deltadefin}
\Delta _{m}^{(0)}(E)=\frac{1}{L}\sum _{\vec{k}\sigma }V_{m\sigma
}(\vec{k})
A_{\sigma }(\vec{k},E)V_{m\sigma }^{*}(\vec{k})
\end{equation}
where the conduction electron spectral function 
\( A_{\sigma }(\vec{k},E)=-\frac{1}{\pi }ImG_{\sigma }(\vec{k},E) \)
depends upon the wave vector \( \vec{k} \) mainly through the band
energy \( \epsilon _{\vec{k}} \) ( see Eq. (\ref{eq:Gkdependence}) ). 
The selfenergy of the empty state is treated in the same manner so the
corresponding selfenergy expressions reduce to 
\begin{eqnarray}
\Sigma _0^{(0)}\left( \omega \right)  & = & {1\over \pi }\sum
_{m}\int dE\, \, 
\Delta _{m}^{(0)}(E)n_f(E)G_{m}\left( \omega +E\right) \nonumber \\
\Sigma _{m}^{(0)}\left( \omega \right)  & = & {1\over \pi }\int dE\,
\, 
\Delta _{m}^{(0)}(E)n_f(-E)G_0\left( \omega -E\right) \label{eq:BubbleSelf} 
\end{eqnarray}
in close analogy to the case of non-interacting electrons \cite{Coleman83}.

The selfconsistency equations Eq. (\ref{eq:BubbleSelf}) were solved 
numerically \cite{TornowDiss} for various well-established 
approximations to the spectral function of interacting conduction 
electrons such as the Hubbard III model \cite{Hubbard64} and the 
Roth approximation \cite{Roth69}. The general results can be summarized
as follows: It is obvious that for (weakly) interacting conduction electrons
the dominant effect of hybridization on the \( 4f^{1} \)
configurational spectrum is a shift 
\( \epsilon _f(U)-\epsilon _f\simeq \Re \Sigma _{m}(\epsilon _f) \)
of the resonance energy which is renormalized by the Coulomb repulsion \( U \)
and its influence on the charge fluctuations. 
The quantity of interest, however,
is the empty-state selfenergy and its variation with energy in the vicinity
of \( \epsilon _f \) which can be deduced from rather simple considerations
assuming that the CEI do not introduce anomalies into the conduction electron
spectral function on the energy scale defined by the characteristic 
temperature
\( T_K \). The smooth variation with energy of 
\( \sum _{\vec{k}}A_{\sigma }(\vec{k},E) \)
implies that in the metallic state the basic analytic structure of 
\( \Sigma _0^{(0)}(\omega ) \)
is not altered as compared to the case of non-interacting 
conduction electrons, the characteristic feature being a 
logarithmic variation in the vicinity of
the f-energy \( \epsilon _f \). The prefactor, however, is proportional to
the interaction-renormalized density of states at the Fermi level 
\( N(\epsilon _f) \).
The low-energy scale, \( T_K \), i.e. the distance between the pole in the
empty f-state Green's function and the \( 4f^{1} \) peak depends on 
the renormalized parameters in the usual exponential way. Especially 
the above is clear for the case when the CEI leads to the spectral 
function of the quasipartical type, \( A_{\sigma }(\vec{k},E) \) 
= \( \delta (E-{\bar{\epsilon }}_{\vec{k}}) \)
with a new dispersion \( {\bar{\epsilon }}_{\vec{k}} \).

Would the single-electron contribution presents the whole story the CEI-case
would be relatively simple. The Coulomb interaction
induces vertex corrections which are of the same order in the inverse 
degeneracy \(1/N \) as the preceding single electron 
contributions. They are an important ingredient of the theory and 
must be included in the discussion 
\cite{Khaliullin95,Tornow97}.

The explicit calculation requires the full Coulomb
vertex \( \Gamma ^{(U)} \) of the conduction electrons as input which must
be determined consistently with the conduction-electron
self-energy.
 We evaluate the 
vertex corrections by analytic continuation from the Matsubara frequencies
inserting the spectral representation 
\begin{equation}
G_{\sigma }(\vec{k},i\omega _{m})=
\int dE\, \frac{A_{\sigma }(\vec{k},E)}{i\omega _{m}-E}
\end{equation}
for the conduction electron propagators and following the rules 
specified in \cite{Bickers87b}. The projection onto
the relevant physical subspace is performed implicitly in the summation over
the Matsubara frequencies where we retain only the contributions from 
the poles in the conduction electron propagators. The empty state self-energy, 
\( \Sigma _0^{(U)}\left( \omega \right)  \),
(see Figure (\ref{fig:BubbleSelf}, a, c))can be written as 
\begin{eqnarray}
\Sigma _0^{(U)}\left( \omega \right)  & = & \frac{1}{\pi }\sum
_{m}\, 
\int dE\, \, \Delta _{0,m}^{(U)}(E,\omega )n_f(E)G_{m}(\omega +E)
\label{eq:Sigma0U} 
\end{eqnarray}
where the Coulomb contribution to the hybridization strength is given by
\end{multicols}
\widetext
\begin{eqnarray}
\Delta _{0,m}^{(U)}\left( E,\omega \right)  & = & \pi {1\over L}
\sum _{\vec{k}\sigma }A_{\sigma }(\vec{k},E)\frac{1}{L^{2}}\sum
_{\vec{k}_1\sigma _1}\! \! \, \, 
\int d\Omega \, \, n_f(-E+\Omega )A_{\sigma
_1}(\vec{k}_1,E-\Omega )
G_0\left( \omega +\Omega \right) \nonumber \\
 &  & \sum _{\vec{k}_2\sigma _2}\! \! \sum _{\vec{k}'\sigma
'_{}}\! \! \, \, 
\sum _{m'}\int dE'\, \, G_{m'}\left( \omega +E'+\Omega _{}\right) \nonumber \\
 &  & \left\{ A_{\sigma _2}(\vec{k}_2,E'+\Omega )n_f(E'+\Omega )
G_{\sigma '}\left( \vec{k}',E'\right) +A_{\sigma
'}(\vec{k}',E')n_f(E')
G_{\sigma _2}\left( \vec{k}_2,E'+\Omega \right) \right\} \nonumber \\
 &  & \left( V_{m\sigma _1}(\vec{k}_1)V_{m'\sigma
_2}(\vec{k}_2)
V_{m\sigma }^{*}(\vec{k})V_{m'\sigma' }(\vec{k}')^{*}
\Gamma _{\sigma _1\sigma _2;\sigma \sigma '}^{(U)}
\left( \vec{k}_1E-\Omega _{},\vec{k}_2E'+\Omega
;\vec{k}_{}E,\vec{k}'E'\right) 
+c.c\right) \nonumber \\
 &  & \delta ^{*}\left(
\vec{k}_1+\vec{k}_2-\vec{k}-\vec{k}'\right) 
\label{eq:Delta0m} 
\end{eqnarray}
\begin{multicols}{2}
\narrowtext
 Here and elsewhere 
\( \Gamma _{\sigma _1\sigma _2;\sigma _3\sigma _4}^{(U)}
\left( 1,2;3,4\right)  \)
is the Coulomb vertex corrections with indices 1,2 for in- and 
3,4 for outgoing particles. 
A similar expression is found for the hybridization strength entering the
occupied-f-states selfenergies, ( see Figure (\ref{fig:BubbleSelf}, b, c )
\end{multicols}
\widetext
\begin{eqnarray}
\Sigma _{m}^{(U)}\left( \omega \right)  & = & \frac{1}{\pi }\int dE\,
\, 
 \Delta _{m,m}^{(U)}(E,\omega )n_f(-E)G_0(\omega -E)
\label{eq:SigmamU} 
\end{eqnarray}
 with 
\begin{eqnarray}
\Delta _{m,m}^{(U)}\left( E,\omega \right)  & = & 
\pi {1\over L}\sum _{\vec{k}\sigma }A_{\sigma
}(\vec{k},E)\frac{1}{L^{2}}
\sum _{\vec{k}_1\sigma _1}\! \! \, \, \sum _{m'}\int d\Omega \, \,
n_f(E-\Omega )
A_{\sigma _1}(\vec{k}_1,E-\Omega )G_{m'}\left( \omega -\Omega \right) 
\nonumber \\
 &  & \sum _{\vec{k}_2\sigma _2}\! \! \sum _{\vec{k}'\sigma
'_{}}\! \! \, \, 
\int dE'\, \, G_0\left( \omega -E'-\Omega \right) \nonumber \\
 &  & \left\{ A_{\sigma _2}(\vec{k}_2,E'+\Omega )n_f(-E'-\Omega
)
G_{\sigma '}\left( \vec{k}',E'\right) +A_{\sigma
'}(\vec{k}',E')n_f(-E')
G_{\sigma _2}\left( \vec{k}_2,E'+\Omega \right) \right\} \nonumber \\
 &  & \left( V_{m\sigma }(\vec{k})V_{m'\sigma' }(\vec{k}')
V_{m\sigma _1}^{*}(\vec{k}_1)V_{m'\sigma _2}^{*}(\vec{k}_2)
\Gamma _{\sigma \sigma ';\sigma _1\sigma _2}^{(U)}
\left( \vec{k}E,\vec{k}'E';\vec{k}_1E-\Omega ,\vec{k}_2E'+\Omega
\right) 
+c.c\right) \nonumber \\
 &  & \delta ^{*}\left(
\vec{k}_1+\vec{k}_2-\vec{k}-\vec{k}'\right) 
\label{eq:Deltamm} 
\end{eqnarray}
\begin{multicols}{2}
\narrowtext
 Note that the self-consistency equations, Eq. (\ref{eq:BubbleSelf})
generalized by including the vertex correction contributions from 
Eqs. (\ref{eq:Delta0m},
\ref{eq:Deltamm}) in the integrands of Eq. (\ref{eq:BubbleSelf}) read
\begin{eqnarray}
\Sigma _0\left( \omega \right)  =  {1\over \pi }\sum _{m}
\int dE\, \, (\Delta _{m}^{(0)}(E) & + &\nonumber\\
\Delta _{0,m}^{(U)}(E,\omega))n_f(E)
G_{m}\left( \omega +E\right) \nonumber \\
\Sigma _{m}\left( \omega \right)   =  {1\over \pi }\int dE\, \, 
(\Delta _{m}^{(0)}(E) & + &\nonumber \\
\Delta _{m,m}^{(U)}(E,\omega))n_f(-E)
G_0\left( \omega -E\right) 
\label{eq:FullSelf} 
\end{eqnarray}

Eqs. (\ref{eq:Delta0m}, \ref{eq:Deltamm}) are general in the sense that they do not 
assume any specific form of conduction electrons spectral functions, vertex corrections,
etc.. In the case when $\vec{k}$- dependences in conduction electron propagators 
enter  as in Eq. (\ref{eq:Gkdependence}) only via the conduction eletrons dispersion 
$\epsilon(\vec{k})$ Eq. (\ref{eq:Delta0m}) may be casted in the form

\begin{eqnarray}   
\Delta _{0,m}^{(U)}(E,\omega)) = \int d\omega_1 d\omega_2 d\omega_3 d\omega_4 \, \, 
\sum_{\sigma_1 \sigma_2 \sigma_3 \sigma_4} \nonumber \\
V_m^{(4)}(\sigma_1 \omega_1, \sigma_2 \omega_2, \sigma_3 \omega_3, \sigma_4 \omega_4)\nonumber\\ 
F_m(E,\omega, \sigma_1 \omega_1, \sigma_2 \omega_2, 
\sigma_3 \omega_3, \sigma_4 \omega_4)
\label{eq:def_VF}
\end{eqnarray}

Here  $V_m^{(4)}$ denotes the 4-order hybridization coupling given
explicitly 
in Eq. (\ref{eq:VFour}) while $ F_m(E,\omega, \sigma_1 \omega_1,
\sigma_2 \omega_2, \sigma_3 \omega_3, \sigma_4 \omega_4)$ is readily obtained by using 
Eq. (\ref{eq:Gkdependence}).  Eq. (22)  
is simplified enormously in the case when conduction electrons 
spectral functions may be approximated by the quasiparticle spectra
. A similar coupling 
$V_m^{(4)}$ may be introduced for the 
$\Delta_{mm}^{(U)}(E,\omega)$. For a model calculation of the 4-order hybridization coupling 
$V_m^{(4)}$ see Appendix \ref{sec:sf}.  

To summarize, we generalized the selfconsistent large degeneracy 
expansion to the case of correlated conduction electrons. The 
modifications due to the interaction enter via the spectral function
of the conduction electrons as well as an effective
renormalized hybridization vertex. The explicit evaluation hence 
requires these quantities for a system of interacting conduction 
electrons. In the subsequent sections, we shall consider the influence 
of the Coulomb repulsion \( U \)
on the effective hybridization strengths which depend upon both \( E \) and
\( \omega  \). In particular, we shall discuss the analytic structure of the
selfenergies for weakly interacting electrons and discuss the
modifications in observable properties in the low-temperature 
limit \( T\rightarrow 0 \).

\section{Weak Conduction-Electron Interaction}

\label{sec:wkceintr}

As a first example, we consider the limit of weakly interacting 
electrons, i.~e.~,
we assume the Coulomb repulsion \( U \) to be much smaller than the bandwidth
\( D \). To leading order in the small ratio \( U/D\ll 1 \) we can neglect
changes in the spectral function of the conduction electrons which we assume
to be given by 
\begin{equation}
\label{eq:ADelta}
A_{\sigma }(\vec{k},E)\longrightarrow \delta (E-\epsilon (\vec{k}))\quad.
\end{equation}
The central focus of the present paper is the lowest pole 
\( \omega _0\) of the empty-f state Green's function 
\begin{equation}
\omega _0-\Sigma _0^{(0)}(\omega _0)-\Sigma _0^{(U)}(\omega
_0)\; =\; 0
\label{eq:pole}
\end{equation} and its variation with the Coulomb repulsion \( U \). 

 The Coulomb repulsion contributes to the configurational selfenergies via the
effective hybridization strengths Eq. (\ref{eq:Delta0m}) and Eq.
(\ref{eq:Deltamm}) where the full Coulomb vertex is replaced by
the bare local non-retarded repulsion ( see Figure \ref{fig:BubbleSelf}, d )

\begin{eqnarray}
\label{eq:lineargamma}
\Gamma _{\sigma _1\sigma _2;\sigma \sigma '}^{(U)}
\left( \vec{k}_1E-\Omega _{},\vec{k}_2E'+\Omega
;\vec{k}E,\vec{k}'E'\right) 
\longrightarrow \\ \nonumber 
U\frac{1}{2}
\left( i\sigma _{y}\right) _{\sigma _1\sigma _2}
\left( i\sigma _{y}\right) _{\sigma \sigma' }
\end{eqnarray}
for the case of $\Delta_{0,m}^{(U)}$ and 
\begin{eqnarray}
\label{eq:lg_m}
\Gamma_{\sigma \sigma ';\sigma_1\sigma_2}^{(U)}
\left( \vec{k}E,\vec{k}'E';\vec{k}_1E-\Omega,\vec{k}_2E'+\Omega \right)
\longrightarrow \\ \nonumber
U \frac{1}{2} \left(i\sigma_y \right)_{\sigma \sigma_1}
 \left(i\sigma_y \right)_{\sigma' \sigma_2} \quad .
\end{eqnarray}
for the case of $\Delta_{m,m}^{(U)}$ accordingly. 

We elaborate on the selfenergies expressions, Eqs. 
(\ref{eq:Sigma0U}, \ref{eq:SigmamU}),
for the case of an {\em orbitally} non-degenerate Anderson model. In
this case the hybridisation matrix element reduces to 
\( V_{m\sigma }(\vec{k})=\delta _{m\sigma }V(\vec{k}) \)
and the occupied f-state propagator does not depend upon the
m-index\cite{CommentOrbitalDegeneracy}. 

Inserting Eqs. (\ref{eq:ADelta}, \ref{eq:lineargamma}, \ref{eq:lg_m})
into the vertex corrections
Eq. (\ref{eq:Delta0m}) and 
Eq. (\ref{eq:Deltamm}) correspondingly
we obtain
\end{multicols}
\widetext
\begin{eqnarray}
\Delta _{0,m}^{(U)}\left( E,\omega \right)  & = & \pi U
\int dE_1\, dE_2\, dE'n_f(-E_1)G_0\left( \omega
+E-E_1\right) 
\frac{2\Re {V^{(4)}(E_1,E_2,E,E')}}{E_1+E_2-E-E'}\nonumber \\
 &  & \left\{ G_{m}\left( \omega +E_2\right) n_f(E_2)-
G_{m}\left( \omega +E'+E-E_1\right) n_f(E')\right\} 
\label{linearDelta_0} 
\end{eqnarray}
and 
\begin{eqnarray}
\Delta _{m,m}^{(U)}\left( E,\omega \right)  & = & \pi U
\int dE'\, dE_3\, dE_4n_f(E_3)G_{-m}\left( \omega
+E_3-E\right) 
\frac{2\Re {V^{(4)}(E,E',E_3,E_4)}}{E_3+E_4-E-E'}\nonumber \\
 &  & \left\{ G_0\left( \omega -E_4\right) n_f(-E_4)-
G_0\left( \omega +E_3-E-E'\right) n_f(-E')\right\} \qquad .
\label{linearDelta_{m}m} 
\end{eqnarray}
\begin{multicols}{2}
\narrowtext
To derive these expressions we used  Eq. (\ref{eq:VFour}),
performed the \( \sigma  \)-summations and the
relevant integrations.

The hybridization matrix elements $V_{\vec{k}}$ vary smoothly 
with $\left|\vec{k}\right|$, and, as a consequence,
$V_m^{(4)}(E_1,E_2,E_3,E_4)$
is a smooth function of the energies $E_i \quad i=1,\dots,4$. It
can be approximated by (see Appendix   \ref{sec:sf})
\begin{equation}
 V_m^{(4)}(E_1,E_2,E_3,E_4) \longrightarrow 
\left(\frac{N(0)\Delta}{\pi}\right)^2 \longrightarrow \left(\frac{\Delta}{2 \pi D}\right)^2
\label{eq:VFoursimple}
\end{equation}
where $\Delta$ is the hybridization width. In the following we shall
adopt a flat density of states extending over the energy range
$(-D,D)$ and use for the $V_m^{(4)}$ the last expression in Eq. (\ref{eq:VFoursimple}).  

We start by discussing the configurational selfenergies for \( T=0 \)
where the Fermi function can be replaced by the step function \(
n_f(x)=\theta (-x) \) and we insert the free propagator for the
occupied-state Green's function \cite{Commentnca}  
\begin{equation}
G_m(\omega) \rightarrow \frac{1}{\omega-\epsilon_f}\quad.
\label{eq:1stiteration}
\end{equation}
The selfenergies
$\Sigma_0^{(U)}$ and $\Sigma_m^{(U)}$ can be expressed in terms of
three integrals $I_{0i}$ and $I_{mi}$, $1=1,2,3$, respectively,
\begin{eqnarray}
\Sigma _0^{(U)}\left( \omega \right)  = 
\frac{1}{2\pi ^{2}}U\left( \frac{\Delta }{D}\right) ^{2} \nonumber \\
\left( -I_{01}(\omega )
\ln \left[ \frac{\omega -\epsilon _f}{\omega -\epsilon
_f-D} \right]  + 
I_{02}(\omega )+I_{03}(\omega ) \right) 
\label{eq:SigmaUBosonTzero} 
\end{eqnarray}
and 
\begin{eqnarray}
\Sigma _{m}^{(U)}\left( \omega \right)   =   -\frac{1}{4\pi ^{2}}U
\left( \frac{\Delta }{D}\right) ^{2} \nonumber \\
\left( -I_{m1}(\omega )\ln \left[\frac{\omega}{\omega-D} \right]  + 
I_{m2}(\omega )+I_{m3}(\omega ) \right) .
\label{eq:SigmaUFermionTzero} 
\end{eqnarray}
Further we discuss the half-filling case\cite{CommentBandFilling}. 
This particular choice
of the band filling, however, does not affect the analytic 
behavior in the energy range of interest, i.~e.~, for $\omega \simeq
\epsilon_f$.

We should like to emphasize that the integrals $I_{0i}$ and $I_{mi}$ in
Eqs. (\ref{eq:SigmaUBosonTzero})
and (\ref{eq:SigmaUFermionTzero})
depend upon the full empty-state Green's function $G_0(\omega)$. This
fact implies that the Coulomb contribution to the selfenergy,
 $\Sigma_0^{(U)}(\omega)$, has to be determined selfconsistently from
Eq. (\ref{eq:SigmaUBosonTzero}) in principle. In the present paper, we 
employ an iterative scheme and adopt a convenient parametrization of
the spectral functions $A_0(\omega)$ and  . Before presenting the results,
let us briefly summarize our procedure. In the first step, we insert
the free empty-state propagator, i.e., $A_0(\omega)\rightarrow\delta(\omega)$ 
into the rhs of Eq. (\ref{eq:SigmaUBosonTzero}). The resulting
selfenergy $\Sigma_0^{(U)}(\omega)$ yields a Green's function
$G_0(\omega)$ which has a Kondo-type pole at
$\omega_{0c}(U)<\omega_0(U=0)<\epsilon_f$ with rather small weight
$1-n_{fc}(U)$. The index c denotes the fact that only the charge
fluctuation contribution was included in the selfenergy
$\Sigma_{0c}^{(U)} (\omega)$. In the next iteration, we account for
the low-energy peak in the spectral function which we model by two
$\delta$-functions $A_0(\omega)\rightarrow
(1-n_{fc}(U))\delta(\omega-\omega_{0c}(U))+n_{fc}(U)\delta(\omega)$.
Including the low-energy spin fluctuations furthers shifts the
threshold $\omega_0(U)$ to lower energy, i.e. we find
$\omega_0(U)< \omega_{0c}(U)<\omega_0(U=0)<\epsilon_f$.

Modelling the
spectral function $A_0(\omega)$ by a combination of $\delta$-functions
allows us to decompose the integrals into contributions from the
charge fluctuations $I_{0ic}(\omega)$, $I_{mic}(\omega)$
( further all paramerers and variables
are in units of the band half-width D )
\begin{eqnarray}
I_{01c}(\omega ) & = &\int _{0}^1\! \! dx \! \! 
\int _0^1\! \! dy \, 
\frac{2\ln \left[\frac{-\omega \! +\! \epsilon _f\! +\! x \! +\! y }
{-\omega \! +\! \epsilon _f\! +\! x \! +\! y \! 
+1}\right]}{(-\omega \! +\! x \! +\! y )(-\omega \! +\!
\epsilon _f\! 
+\! x )}\label{eq:I01c} \\[3mm]
I_{02c}(\omega ) & = & \left(\int _{0}^1\! \! dx \! \! \,
\frac{\ln\left[\frac{-\omega+x+1}{-\omega + x}\right]}{(-\omega \! +\! \epsilon _f+\! x )}\right)^2
\label{eq:I02c} \\[3mm]
I_{03c}(\omega ) & = & \int _{0}^1\! \! dx \! \! \int
_0^{1}\! \! dy \! \! 
\int _{0}^1\! \! dz 2\ln\left[ 
\frac{x +y +z +1}{x +y +z }\right] \nonumber \\
 & \times & \left[(-\omega \! +\! x \! +\! y)
(-\omega \! +\! \epsilon _f\! +\! x ) \right. \nonumber \\
 & \times & \left.(-\omega \! +\! \epsilon _f\! +\! x \! +\! y \! +\! z )\right]^{-1}
\label{eq:I03c} \\[3mm]
 & & \nonumber\\[3mm]
I_{m1c}(\omega ) & = & \int _{0}^1\! \! dx \! \! 
\int _{0}^1\! \! dy \!
\frac{2\ln \left[\frac{-\omega+x+y}{-\omega+x+y+1}\right]}{(-\omega \! +\! x )
(-\omega \! +\!\epsilon_f +\! x \! +\! y )}
\label{eq:Im1c} \\[3mm]
I_{m2c}(\omega ) & = & \left(\int _{0}^1\! \! dx \! \! \,
\frac{\ln \left[\frac{-\omega+\epsilon_f+x}{-\omega+1+\epsilon_f+x}\right]}
{(-\omega \! +\! x)}\right)^2
\label{eq:Im2c} \\[3mm]
I_{m3c}(\omega ) & = & \int _{0}^1\! \! dx \! \! \int
_{0}^{1}\! \! dy \! \! 
\int _{0}^1\! \! dz 
2\ln\left[\frac{x+y+z+1}{x+y+z }\right] \nonumber \\
 & \times & \left[(-\omega \! +\! x)
(-\omega \! +\! \epsilon _f\! +\! x \!+\! y) \right. \nonumber \\
& \times & \left.(-\omega \! + \! x \! +\! y \! + \! z )\right]^{-1}
\label{eq:Im3c}
\end{eqnarray} 
and from spin fluctuations integrals $I_{0isf}$ and  $I_{misf}$. The latter integrals are obtained from 
their charge fluctuations counterparts by the substitution $\omega \rightarrow \omega-\omega_0$. 
The charge fluctuations integrals have no singularities for $\omega < \epsilon_f < 0$ and it is evident just 
from their inspection that $I_{01c},I_{m1c} < 0$ and other integrals are positive.

\begin{figure}[bt]
\begin{center}
\includegraphics[width=8.0cm]{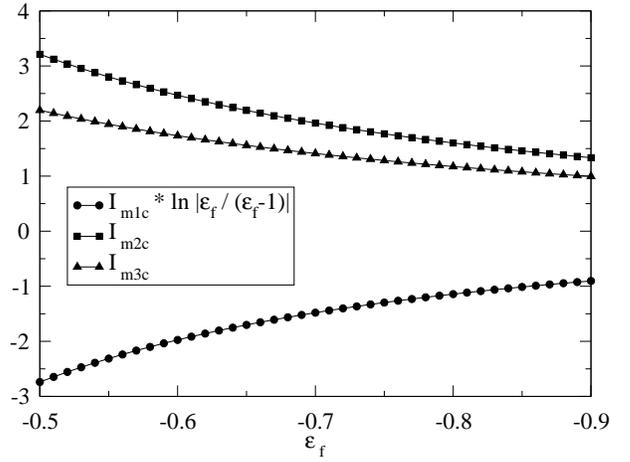}
\end{center}
\vspace{10pt}
\caption{Contributions to the occupied-state selfenergy from the
Coulomb correction to the effective hybridization vertex:
Occupied-state integrals $I_{mic}(\epsilon_f)$ evaluated for various
values of the f-level energy $\epsilon_f$. Here and in all further Figures solid lines 
are guides for the eye. }
\label{fig:Imic}
\end{figure}
The spin fluctuations 
integrals are of analogical properties
 but for $\omega < \omega_0$. The 
infinitesimal imaginary parts $+i0_{+}$   in denominators of the integrands 
in Eqs. (\ref{eq:I01c}-\ref{eq:Im3c}) are
omitted because they do not contribute for $\omega < \epsilon_f$.  
Note that for $n_f < 1$  integrals
$I_{0ic}$, $I_{0is}$ have to be inserted in Eq.(\ref{eq:SigmaUBosonTzero}) being multiplied by $n_f$ or 
$1-n_f$ correspondingly.
The contributions from the spin fluctuations to the Coulomb
renormalization of the occupied-state vertex $I_{misf}$ are neglected.
The Coulomb contributions to the occupied-state selfenergy
vary rather smoothly with $\omega$ in the vicinity of $\epsilon_f$.
\begin{figure}[t]
\begin{center}
\includegraphics[width=8.0cm] {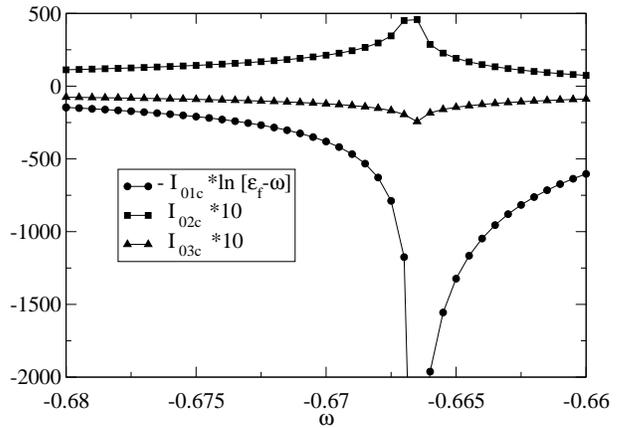}
\end{center}
\vspace{10pt}
\caption{ Charge fluctuation contribution to
empty-state selfenergy integrals $I_{0ic}(\omega)$ for
$\epsilon_f=-2/3$, $\Delta=0.2$, $U=0.1$ For $\omega > \epsilon_f$
the real part is shown.}
\label{fig:I0ic}
\end{figure}
They
give rise to a rather small shift of the effective f-level which can be
estimated from the integrals $I_{mic}(\epsilon_f)$ displayed in Figure
 \ref{fig:Imic} and the
real part of the selfenergy $\Sigma_m^{(U)}(\epsilon_f)$. As we shall see below, 
we need not explicitly
account for the shift in the determination of the many-body low-energy scale.

Let us now turn to the empty state selfenergy. Following the iterative 
procedure we include the charge fluctuations in the first step.
The variation with energy of the integrals $I_{0ic}(\omega)$ is
displayed in Figure \ref{fig:I0ic}.

A detailed analysis shows that in the energy range of interest
the integrals vary approximately like 
\( I_{01c}(\omega )\sim A_1\ln |\omega -\epsilon _f|+B_1 \)
and \( I_{02c}(\omega )\sim (A_2\ln |\omega -\epsilon _f|+B_2)^{2} \)
where $A_1 > A_2$. The resulting real part of the empty-f state
self-energy  varies like 
$ -(\log|\epsilon_f-\omega |)^2 $ in the
vicinity of the (renormalized) f-level. 
As a consequence, the corresponding Green's 
function $G_0(\omega)$ always exhibits a pole at $\omega_0(U) <
\omega_0(U=0)<\epsilon_f$.
For not too small values of the hybridization width
they are well described by the linear dependence
\begin{equation}
\omega_0 (U) \simeq \omega_0(U=0) +\frac{\displaystyle \Re
\Sigma_{0c}^{(U)}(\omega_0)}{\displaystyle 1-
\left[\frac{\partial\Re\Sigma_0^{(0)}(\omega)}{\partial
\omega}\right]}_{\omega_0}
\label{eq:LinearInUOmega0}
\end{equation}
The change in the pole is seen to be proportional to the weight of the
f$^0$-configuration in the ground state times the Coulomb repulsion
among the conduction electrons.

For a first qualitative  understanding of the variation with the
Coulomb repulsion of the threshold energy $\omega_0$ 
 one may use ''on-shell'' approximation \cite{Tornow97}. Within this approximation, \marginpar{!}
the empty-state selfenergy $\Sigma_0(\omega)=\frac{2{\tilde \Delta}}{\pi}
\ln \left[{\tilde \epsilon}_f-\omega\right]$ has the same $\omega$- dependence as in the 
non-interacting case but with renormalized parameters: $ {\tilde \Delta}  = 
\Delta \left( 1-\frac{U}{4\pi }\Delta I_{01c}
\left( \omega _0\right) \right); 
{\tilde \epsilon}_f  =  \epsilon _f-\frac{U \Delta^2}{2\pi ^{2}}
\left(I_{02c}\left( \omega _0\right) +
 I_{03c}\left( \omega _0\right)\right)$
with $\omega_0 = \omega_0(U=0)$ here. We see that  ${\tilde \Delta} > \Delta$ and 
$|{\tilde \epsilon}_f| > |\epsilon_f|$. 
If the impurity valence is close to integer  
which lead to a Kondo regime the renormalization of the hybridization coupling 
prevails the renormalization 
of the f-level energy resulting in an effective enhancement of the
Kondo energy scale. The simplified approach, however, cannot be used
for quantitative estimates. 
Unfortunately the variation with $U$ of the corresponding Kondo-type pole
is systematically underestimated ( $T_K$ is overestimated ) as can be easily seen from the slopes 

$\frac{\partial \omega_0^{OnShell}}{\partial U}-
\frac{\partial \omega_0}{\partial U}=
-\frac{1}{2\pi^2}\Delta^2
\left(I_{02c}(\omega_0)+I_{03c}(\omega_0)\right) < 0$

\begin{figure}[bt]
\begin{center}
\includegraphics[width=8.0cm]{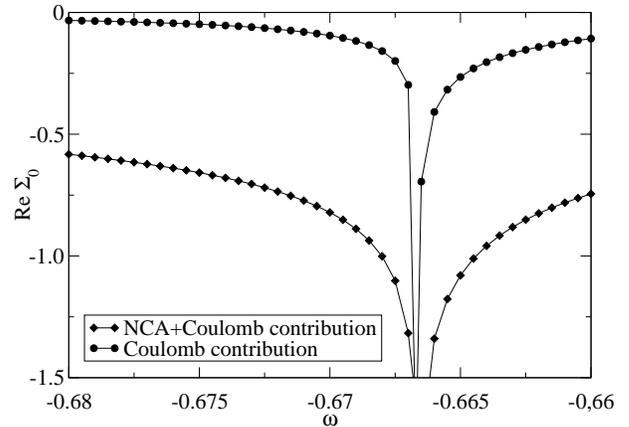}
\end{center}
\vspace{10pt}
\caption{Variation with energy of the real part of the full empty-state 
selfenergy $\Re \Sigma_0^{(U)}(\omega)$ from NCA plus Coulomb correction
to the hybridization vertex for $\epsilon_f=-2/3$, $\Delta=0.2$ and $U=0.1$}
\label{fig:Sigma0}
\end{figure}

The results for the Kondo pole change only slightly upon
iteration. Inclusion of the spin fluctuation contribution to the
Coulomb correction yields a rather small shift in the selfenergy which 
further stabilizes the Kondo ground state. This can be seen from the calculated \achtung
variation with energy of the integrals $I_{0isf}(\omega)$. The full selfenergy $\Sigma_0(\omega)$
is shown in Figure  \ref{fig:Sigma0}.
The characteristic energy scale for low-energy excitations, i. e., the 
Kondo temperature, is now calculated as the difference between the
ground state energy - the threshold $\omega_0$ - and the energy of the 
f-level $\epsilon_f$
\begin{equation}
k_BT_K = \epsilon_f - \omega_0
\label{eq:TKondoDef}
\end{equation}
At this point we should like to add a comment concerning the choice of 
 $\epsilon_f$. This quantity enters Eq. (\ref{eq:TKondoDef}) explicitly
as well as implicitly through $\omega_0$. If we were to account for
the Coulomb renormalization we would have to do it consistently. This
means we would have to consider the difference  \marginpar{!}

$\epsilon_f+\delta \epsilon_f  - \omega_0(\epsilon_f+\delta
\epsilon_f) \simeq 
\epsilon_f - \omega_0 +\delta \epsilon_f \left(1-n_f\right)$.
The correction from the Coulomb contribution is hence proportional to
$\delta \epsilon_f \left(1-n_f\right) $ which is rather small in our case 
because the shift $\Re\Sigma_m(\omega)$ \achtung
is very small. 
\begin{figure}[bt]
\begin{center}
\includegraphics[width=8.0cm]{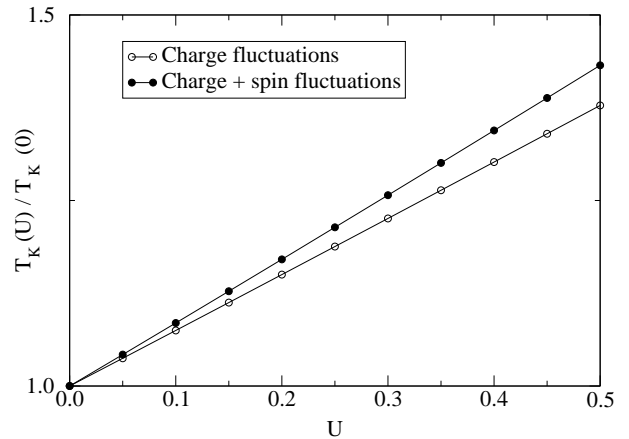}
\end{center}
\vspace{10pt}
\caption{Enhancement of the Kondo temperature for $\epsilon_f=-2/3$
and $\Delta=0.2$. Charge fluctuations are dominant. }
\label{fig:TKondoChargeVsSpin}
\end{figure}
 
To conclude, 
we can calculate $k_BT_K$ from Eq. (\ref{eq:TKondoDef}) using the
the bare f-level energy at $\epsilon_f$.
The results displayed in Figure \ref{fig:TKondoChargeVsSpin} 
 show the enhancement of the Kondo temperature
due to the Coulomb interaction among the conduction electrons. 

Following up the iteration procedure the spectral function
\begin{equation}
A_0(\omega)=-\frac{1}{\pi} \Im
\left(\omega-\Sigma_0^{(0)}-\Sigma_{0c}^{(U)}\right)^{-1}
\end{equation}
is inserted in to Eq. (\ref{eq:SigmaUBosonTzero}). It comes from calcultions \achtung
 that including the spin
fluctuation peak at $\omega=\omega_0(U)$ does not significantly alter
the empty-state selfenergy $\Sigma_0^{(U)}(\omega)$ in the Kondo
regime.  The spin fluctuations lower the energy of 
the pole in the Green's function $G_0(\omega)$ and therefore further 
stabilize the Kondo ground
state. This can be seen from Figure \ref{fig:TKondoChargeVsSpin}. The 
data suggest the charge fluctuation contribution already gives 
a rather good estimate
of the Coulomb correction to the low-temperature properties to leading 
order in the inverse degeneracy.

The main feature of the above caculations
is the factorization of the `NCA-bubble' self-energy \( \Sigma _0^0(\omega ) \)
in the r.h.s. of Eq. (\ref{eq:SigmaUBosonTzero}) for the empty state self-energy
\( \Sigma _0^{(U)}(\omega ) \). This factorisation is due to the possibility,
as it was shown for the orbital degeneracy case in Appendix \ref{sec:sf},
to neglect the momentum conservation in the integrals of the $V^{(4)}$ coupling, Eq. 
(\ref{eq:VFour}).
Therefore results of this section are also
valid for the degenerate case if one replace in Eq. (\ref{eq:SigmaUBosonTzero})
\( \Delta \rightarrow (N-1)\Delta  \). So the renormalisation of the paramerers
of the Anderson impurity Hamiltonian is selfconsistent, in the spirit of the
NCA. 


\section{Thermodynamic properties at Zero Temperature}
\label{sec:thermod}
The results of the preceeding section allow us to assess the influence 
of the conduction electron Coulomb repulsion on the thermodynamic
properties of dilute magnetic alloys. Of particular interest are the
low-temperature f-valence $n_f$, the $f$-charge susceptibility
$\chi_c$, the $f$-spin susceptibility $\chi_s$ and the magnetic
contribution to the linear coefficient of the specific heat $\gamma$.
Previous calculations based on the symmetric Anderson model yield a
rather strong depression with U of the $f$ magnetic susceptibility
\cite{Hofstetter00}. Data for the U-dependence of the f valence $n_f$
and the f charge susceptibility $\chi_c$, however could not be
obtained from these model studies since particle-hole symmetry pins
$n_f$ to unity. For a first quantitative estimate we approximate the
empty-state selfenergy $\Sigma_0$ by
\begin{equation}
\Sigma_0^{(0)} \rightarrow \Sigma_0^{(0)}+\Sigma_{0c}^{(U)}
\end{equation}
keeping only the charge fluctuation contribution. This procedure should
be justified in the Kondo limit where the deviation from integer f
valence is small. The pole of the corresponding Green's function,
$\omega_0 $, can be interpreted as the ground state of the system. It
yields the dominant low-temperature
contribution to the partition function, and the 
thermodynamic properties follow by straightforward differentiation \cite{Hewsonbook}
\begin{eqnarray}
n_f & = & \frac{\partial \omega_0}{\partial \epsilon_f}
\label{eq:fValence}  \\
\chi_c & = & - \frac{\partial^2 \omega_0}{\partial \epsilon_f^2} 
\label{eq:ChiCharge}\\
\chi & = & \lim_{H\to0}
\left(- \frac{\partial^2 \omega_0}{\partial H^2}\right)
\label{eq:ChiSpin} \\
\gamma & = &  \lim_{T\to0}
\left(- \frac{1}{T}\frac{\partial \omega_0(T)}{\partial
T}\right)\label{eq:Gamma} 
\end{eqnarray}
Finally, we shall discuss the Sommerfeld-Wilson ratio $R$, i.~e.~, the
ratio of the zero-temperature spin susceptibility and the specific
heat coefficient
\begin{equation}
R=\frac{\displaystyle \frac{\pi^2}{3} \chi}{\displaystyle
\frac{\mu_j^2}{3} \gamma}
\label{eq:SommerfeldWilson}
\end{equation}
where $\mu_j^2=j(j+1)(g\mu_B)^2$.

Our main interest is in the linear in U corrections to the
experimental quantities. These contributions can be easily obtained
from the linear in U corrections to the ground state energy as given
by Eq. (\ref{eq:LinearInUOmega0}). 
We specify the interaction related enhancement/reduction in terms of the
coefficients
\begin{eqnarray}
n_f(U) & \simeq & n_f^{(0)}\left(1+U n_f^{(1)} \right) 
\nonumber \\[2mm]
\chi_c(U) & \simeq & \chi_c^{(0)}\left(1+U \chi_c^{(1)} \right) 
\nonumber \\[2mm]
\chi_s(U) & \simeq & \chi_s^{(0)}\left(1+U \chi_s^{(1)} \right)
\nonumber \\[2mm]
\gamma_f(U) & \simeq & \gamma_f^{(0)}\left(1+U \gamma_f^{(1)}\right)
\nonumber  \\[2mm]
R(U) & \simeq & R^{(0)}\left(1+U R^{(1)} \right)
\label{eq:Reduced}
\end{eqnarray}
which depend upon the f-level position $\epsilon_f$ and the
hybridization width $\Delta$ and, concomitantly, on the Kondo energy
$k_BT_K^{(0)}$  of the reference system with non-interacting
conduction electrons.

The explicit evaluation requires the generalization of
$\Sigma_0^{(U)}$ to low but finite temperatures and to small external
magnetic fields. The former is easily achieved by starting from 
Eq. (\ref{linearDelta_0}) and proceeding in
close analogy to the zero temperature case keeping the Fermi functions
instead of the step functions. The derivatives with respect to
temperature are calculated from a Sommerfeld expansion. An external
magnetic field, on the other hand, lifts the 
the degeneracy of the
f-level according to \cite{CommentSigmah} 
\begin{equation}
\epsilon_f \longrightarrow \epsilon_f +\sigma h \quad .
\label{eq:efh}
\end{equation}
The NCA contribution $\Sigma_0(\omega)=\frac{\Delta}{\pi} \sum_\sigma
\ln \left[\epsilon_f+\sigma h -\omega \right] $ is directly read
off. The Coulomb contribution 
is now expressed in terms of three spin-dependent integrals 
$I_{0i}(\omega;\sigma)$
which which closely parallel their counterparts in the absence of an 
external magnetic field $I_{0i}(\omega)$. Keeping only the charge 
fluctuation contribution yields
\end{multicols}
\widetext
\begin{eqnarray}
\Sigma _0^{(U)}\left( \omega \right) & = &
\sum_{\sigma}
\frac{1}{{\bf 4\pi ^{2}}}U \Delta^{2}
\left( -I_{01}(\omega; \sigma)
\ln \left|{ \epsilon _f-\sigma h-\omega}\right|+ I_{02}(\omega;\sigma
)+I_{03}(\omega;\sigma )\right) \\[3mm]
I_{01c}(\omega;\sigma ) & = &\int _{0}^1\! \! dx \! \! 
\int _0^1\! \! dy \, 
\frac{2}{(-\omega \! +\! x \! +\! y )(-\omega \! +\!
\epsilon _f\! +\sigma h\!
+\! x )}\ln \left[\frac{-\omega \! +\! \epsilon _f\! -\sigma h\! +\! x \! +\! y }
{-\omega \! +\! \epsilon _f\! -\sigma h\!+\! x \! +\! y \! 
+1}\right]\label{eq:I01ch} \\[3mm]
I_{02c}(\omega;\sigma ) & = & \tilde{I}_{02c}(\omega;\sigma)\tilde{I}(\omega;-\sigma)\nonumber\\
\tilde{I}_{02c}(\omega;\sigma ) & = & \int _{0}^1\! \! dx \! \! \,
\frac{1}{(-\omega \! +\! \epsilon _f +\!\sigma h+\! x )}\ln\left[\frac{-\omega+1+x}{-\omega+x}\right]
\label{eq:I02ch} \\[3mm]
I_{03c}(\omega;\sigma ) & = & \int _{0}^1\! \! dx \! \! \int
_0^{1}\! \! dy \! \! 
\int _{0}^1\! \! dz \frac{2 \ln\left[ 
\frac{x +y +z +1}{x +y +z }\right]}{(-\omega \! +\! x \! +\! y
)(-\omega \! +\! \epsilon _f\! +\! \sigma h \! +\! x )
(-\omega \! +\! \epsilon _f\! -\! \sigma h \!+\! x \! +\! y \! +\! z
)} 
\quad .
\label{eq:I03ch} 
\end{eqnarray}
\begin{multicols}{2}
\narrowtext

\begin{figure}[bt]
\begin{center}
\includegraphics[width=8.0cm]{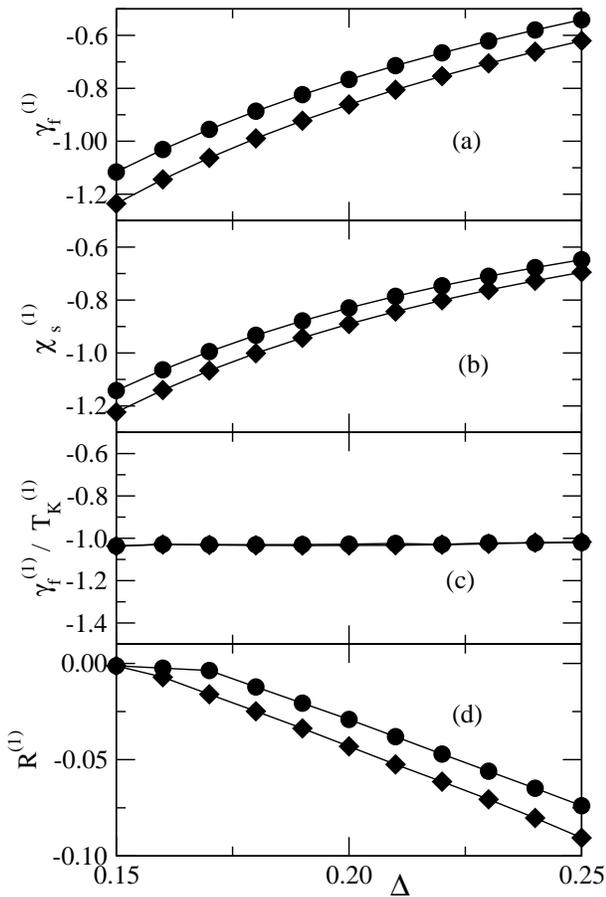}
\end{center}
\vspace{10pt}
\caption{Reduced thermodynamical coefficients from Eq. \ref{eq:Reduced} as functions of $\Delta$
for $\epsilon_f=-2/3$ (diamonds) and $\epsilon_f = - 0.7$ (circles).
(a):${\bar \gamma}_f^{(1)}$, (b):${\bar \chi}_s^{(1)}$, (c):${\bar
\gamma}_f^{(1)} {\bar T_K^{(1)}}$, (d):${\bar R}^{(1)}$}
\label{fig:GammaChis}
\end{figure}
At low temperatures, we find a finite temperature-independent
Pauli-like spin susceptibility and a linear specific heat indicating a
nonmagnetic Fermi liquid ground state.
The results are displayed in Figure \ref{fig:GammaChis}(a) and (b). The
coefficient $\gamma_f$ of the linear specific heat is reduced by the
conduction electron interactions reflecting the enhancement of the
Kondo temperature. The scaling $\gamma_f \sim 1/T_K$ can be seen from
from Figure \ref{fig:GammaChis}(c)\cite{$T_K^{(1)}$}. Similarly, the magnetic
susceptibility is reduced by the conduction electron interactions. 
The reduction determined here is comparable to the value
obtained by Hofstetter et al. \cite{Hofstetter00}. Its 
actual values, however, exhibit deviations from universal scaling with the
inverse Kondo temperature $1/T_K$ reflecting the importance of
quasiparticle interactions. This is to be expected from the
explicit expression for the spin susceptibility calculated to 
leading order in the conduction electron interaction

\begin{eqnarray}
 \chi_s(U)/\chi_s^{(0)}  \simeq  1 
-  \left[\frac{\partial}{\partial \omega} \log \left(-\frac{%
\frac{\partial^2 \Sigma_0^{(0)}}{\partial h^2}}{%
1-\frac{\partial \Sigma_0^{(0)}}{\partial
\omega}}\right)\right]_{\omega_0^{(0)},h=0}
\delta T_K \nonumber\\
  +  
\left(\left[\frac{%
\frac{\partial \Sigma_0^{(U)}}{\partial \omega}}{%
1-\frac{\partial \Sigma_0^{(0)}}{\partial
\omega}}\right]_{\omega_0^{(0)},h=0}+
\left[\frac{%
\frac{\partial^2 \Sigma_0^{(U)}}{\partial h^2}}{%
\frac{\partial^2 \Sigma_0^{(0)}}{\partial
h^2}}\right]_{\omega_0^{(0)},h=0}\right)\quad 
\end{eqnarray}


The interaction correction to the spin susceptibility consists of 
two terms where the first one is proportional to the change in the Kondo
temperature. As its coefficient varies proportional to $T_K (0)^{-2}$ 
we expect this contribution to dominate in the close to integer
valence limit where $T_K(0)$ becomes small. In this limit we should
recover the typical Kondo scaling. The second
term which results from the variation with h of the interaction
corrections tends to further reduce the spin susceptibility. Within
the spirit of Landau's Fermi liquid theory it gives rise to a positive 
Landau parameter $F_0^{(a)}$.

The deviation from simple scaling is also reflected in the
Sommerfeld-Wilson ratio which is reduced by the conduction electron 
Coulomb interaction (see Figure \ref{fig:GammaChis}(d)). Since
\begin{equation}
 {\bar R}^{(1)} \simeq  {\bar \chi}_s^{(1)} -  {\bar \gamma}_f^{(1)}.
\label{eq:R_1}
\end{equation}
this quantity allows us to estimate the Landau parameter
$ {\bar R}^{(1)} \simeq - F_0^{(a)}$.

\bigskip

\begin{figure}[bt]
\begin{center}
\includegraphics[width=8.0cm]{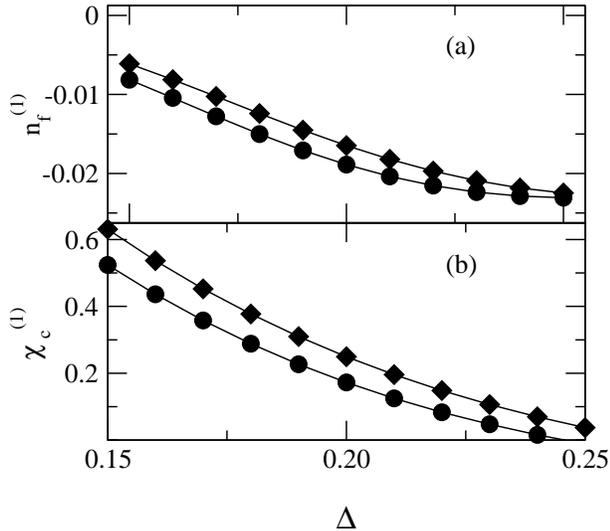}
\end{center}
\vspace{10pt}
\caption{Reduced thermodinamical coefficients from Eq. \ref{eq:Reduced} as functions of $\Delta$
for $\epsilon_f=-2/3$ (diamonds) and $\epsilon_f = - 0.7$ (circles).
(a):${\bar n}_f^{(1)}$, (b):${\bar \chi}_c^{(1)}$}
\label{fig:nfChic}
\end{figure}
The deviations from the universal scaling with $T_K$ are most strongly 
evident in the f-valence, $n_f$, and in the f-charge susceptibility,
$\chi_c$ displayed in Figure \ref{fig:nfChic}. The f-valence is slightly decreased
by the conduction electron Coulomb repulsion as can be seen from the
coefficient $ {\bar n}_f^{(1)}$ displayed in Figure
\ref{fig:nfChic}, (a).
 This behavior reflects two competing effects, i. e., (a) the increase in characteristic
energy as demonstrated in Figure \ref{fig:TKondoChargeVsSpin} 
 which is partially compensated by (b) the
enhancement of the effective hybridization. 

The f-charge susceptibility $\chi_c$ is affected more dramatically as
suggested by the coefficient ${\bar \chi}_c^{(1)}$ in Figure
\ref{fig:nfChic}, (b). In the Kondo regime,  it is 
enhanced by the Coulomb interaction of the
conduction electrons. This enhancement, however, decreases as we
approach the mixed-valent regime. In this parameter regime, however,
the adopted approximation (charge fluctuation contribution to
selfenergy only) ceases to be valid.

To summarize, the low-temperature properties of a dilute magnetic
alloy are significantly modified by the Coulomb repulsion of
the conduction electrons. The local Fermi liquid properties
are preserved in the sense that the magnetic contribution to the
specific heat varies linearly with temperature, $C_f \sim T$, and that 
the magnetic susceptibility $\chi_s$ is finite at $T = 0$. 
Even in the lowest order in the
inverse degeneracy, the low-temperature properties are not universal
in the sense that their variation with U cannot be accounted for by
properly adjusting the Kondo temperature. 


\section{Discussions}

\label{sec:dscssns}

We formulated here the NCA equations for an Anderson impurity coupled
to interacting conduction electrons. Due to the CEI an
energy-dependent effective
hybridization vertex appears in the usual NCA equations.
The case of weak CEI  was investigated in detail here including calculations of 
measurable thermodynamical properties of a dilute magnetic impurities system. 
The results for a rigid, i. e., energy-independent vertex can serve as
a guide-line for the case of a magnetic impurity embedded in a Fermi
liquid where the central quantity is the quasiparticle t-matrix 
\cite{CommentResidIntrctn}. 
 
For weak electron-electron interaction the increase of \( T_K \) may be understood
as resulting from the reduced probability of finding doubly occupied and empty
lattice sites in the correlated conduction electron systems. The increased number
of uncompensated conduction electron spins finally leads to the enhancement
of the effective hybridization coupling. The analogy between the Kondo spin
model and the Anderson impurity model in its local moment regime is not complete
in the case of correlated conduction electrons ( see also \cite{Khaliullin95}
). In the former case \( T_K \) will increase monotonously with \( U \)
because of the enhancement of the exchange interaction. In the latter case the
process is two-staged, it involves the formation of a local moment and its interaction
with the conduction electrons. As we show here CEI influence thermodynamical quantities 
in a non-trivial fashion: the specific heat coefficient $\gamma(U)$ scales with $T_K(U)$ 
like in the non-interacting case while the magnetic susceptibility $\chi_s(U)$ does not.
That may be understood because the latter quantity depends not only on the low energy 
excitations spectrum as the former but also on the matrix element which is influenced by CEI. 

There is some controversy about the scaling with
$T_K(U) $ of $\chi_s(U)$. In a somewhat  
 artificial model of \cite{Takayama98} the scaling of $\chi_s(U)$ with $T_K(U)$ is preserved 
for not too large values of $U$. In the paper \cite{Hofstetter00} this question is not discussed 
explicitly but it may be judged from the relevant plots in
\cite{Hofstetter00} that their results do not
show the usual scaling behaviour of $\chi_s(U)$ but rather resemble our results. 

The present results are based on the separation of energy scales. In so far
as the vertex correction does not produce a new low frequency scale or does
not dominatly contribute to the empry state ( empty-f-state ) self-energy we
anticipate no qualitative changes when using more sophisticated approximations
for the vertices \( \Gamma _{\sigma ,\bar{\sigma }}^{\left( U\right) }\left( 1,2;3,4\right)  \),
(Figure \ref{fig:BubbleSelf}\ c) appropriate for the strong correlation regime.
We may expect that for sufficiently large \( U \) the virtual transition from
the f-state to the conduction state will cost too much energy inhibiting the
\( T_K \)-increase and leading eventually to the change in the trend 
\cite{Davidovich98,Hofstetter00}. For
a quantitative treatment of this problem there are two visible approaches. One
 is to introduce summation of the infinite subseries of the ring, the ladder
and the particle-particle type starting with the 2nd order
self-energies \cite{TornowDiss}. 
The other way is to approximate the vertex correction by response
functions ( dynamic susceptibilities ). 
The other visible
development of the theory of the Kondo effect for electronic correlated system
in general and high-\( T_{c} \) cuprates in particular is the use of the antiferromagnon
dynamic susceptibility \cite{Chubukov02} or other phenomenological models which 
develop a short-range order with the virtual breaking of singlet-triplet degeneracy
in the conduction-electron system ( see also Ref. \cite{Davidovich98} ). In addition,
the influence of conduction electron interactions on the spectral properties
of magnetic impurities and their dependence upon the doping are also interesting
topics for future investigations.

In conclusion, the NCA theory of an Anderson impurity embedded in a metall with
correlated conduction electrons is developed and general NCA equations for the
interacting conduction electrons are obtained. It is shown that due to the renormalisation
of the hybridization interaction the characteristic energy is increased by the
weak interactions. The influence of weak conduction electron interations on thermodynamic 
properties of magnetic impurities is discussed. 

\section{Acknowledgments}
 
The hospitality at TU-Braunschweig is acknowledged by V.Z.
This work was supported by the Nieders{\"a}chsisch-Israelischer Foundation.
Discussions with Dr. A. Schiller are appreciated.
\appendix
\section{Fourth order term} 
\label{sec:sf} 
We estimate the variation with
the orbital degeneracy \( N \) of the vertex-corrected boson selfenergy (Figure
\ref{fig:BubbleSelf}(c) assuming a simple band structure for the metallic host.
The orbitally degenerate conduction bands are modelled by tight-binding s-bands
the dispersion being determined by hoping between nearest neighbor sites on
a simple cubic lattice. The magnetic impurity sitting at the origin is surrounded
by six nearest neighbors. The wave vector dependence of the coupling between
the conduction states and the strongly correlated local orbital \( (j,m) \)
is given by the canonical structure constant \( S_{00;jm}(\vec{k}) \) \cite{Andersen75,SkriverBook}
of the Atomic Sphere Approximation (ASA) to the Linear Muffin Tin Orbital (LMTO)
method. Here \( j \) and \( m \) refer to the azimuthal and magnetic quantum
number of the impurity orbital under consideration.

Let us neglect spin-orbit interaction for a first qualitative discussion. The
hybridization matrix element \( V_{m\sigma }(\vec{k}) \) does not depend upon
\( \sigma  \). It is given by 
\begin{equation}
\label{eq:hybasa}
V_{m\sigma }(\vec{k})=V_0({\epsilon _{\vec{k}}})\sum _{\vec{R}_{j}}e^{i\vec{k}\cdot \vec{R}_{j}}\left[ \sqrt{4\pi }iY_{\ell m}(\hat{R}_{j})\right] ^{*}\left( \frac{S}{\left| \vec{R}_{j}\right| }\right) ^{\ell +1}
\end{equation}
 where the argument \( \hat{R}_{j} \) of the spherical harmonic \( Y_{\ell m} \)
denotes the unit vector pointing from the impurity to the nearest neighbor sites
\( \vec{R}_{j} \). The overall length scale \( S \) is usually chosen as the
average atomic radius while \( V_0(\epsilon _{\vec{k}}) \) is an energy-dependent
real prefactor.

Starting from this form of the hybridization we shall first derive an expression
for the second order term which is subsequently compared to the fourth order
average.

The sum 
\begin{equation}
\sum _{m}V_{m\sigma }^{*}(\vec{k})V_{m\sigma }(\vec{k}')
\end{equation}
 is easily evaluated using the addition theorem for spherical harmonics 
\begin{eqnarray}
\sum _{m_{-}\ell }^{\ell }\left( Y_{\ell m}(\hat{R}_{j})\right) ^{*}Y_{\ell m}(\hat{R}_{j'})=
\frac{2\ell +1}{4\pi }P_{\ell }\left( \hat{R}_{j}\cdot \hat{R}_{j'}\right) = \nonumber \\
\frac{N}{4\pi }P_{\ell }\left( \hat{R}_{j}\cdot \hat{R}_{j'}\right) 
\end{eqnarray}
 where \( N=2\ell +1 \) is the degeneracy of the impurity level. For a simple
cubic lattice with 
\begin{equation}
\hat{R}_{j}\cdot \hat{R}_{j'}=\delta _{\hat{R}_{j}\hat{R}_{j'}}-\delta _{\hat{R}_{j},-\hat{R}_{j'}}
\end{equation}
 we obtain for \( \ell =3 \) 
\begin{eqnarray}
\sum _{m}V_{m\sigma }^{*}(\vec{k})V_{m\sigma }(\vec{k}')V_0(\epsilon _{\vec{k}})V_0(\epsilon _{\vec{k}'})N \times
\nonumber \\
\sum \, \, _{\vec{R}_{j}}\left( e^{-i\vec{k}\cdot \vec{R}_{j}}e^{i\vec{k}'\cdot \vec{R}_{j}}-
e^{-i\vec{k}\cdot \vec{R}_{j}}e^{-i\vec{k}'\cdot \vec{R}_{j}}\right) \quad .
\end{eqnarray}
 As expected, the second order term entering the non-crossing diagram 
\begin{eqnarray}
\langle \sum _{m}\left| V_{m\sigma }(\vec{k})\right| ^{2}\rangle _{\epsilon } & = & \left| V_0(\epsilon )\right| ^{2}N\langle \sum \, \, _{\vec{R}_{j}}\left( 1-\cos 2\vec{k}\cdot \vec{R}_{j}\right) \rangle _{\epsilon }\nonumber \\
 & = & \left| V_0(\epsilon )\right| ^{2}N2\langle \sum \, \, _{\vec{R}_{j}}\sin ^{2}\vec{k}\cdot \vec{R}_{j}\rangle _{\epsilon }\nonumber \\
 & = & \left| V_0(\epsilon )\right| ^{2}2zN\langle \sin ^{2}k_{x}\rangle _{\epsilon }
\end{eqnarray}
 is proportional to the degeneracy \( N=2\ell +1 \) and the number of nearest
neighbbors with \( z=6 \) for s.c.l.. The average over the constant energy
surface which has to be numerically evaluated is of order unity and varies smoothly
with the energy \( \epsilon  \). To summarize: The non-crossing diagrams involve
the combination 
\begin{equation}
N\Delta (\epsilon )=NN(\epsilon )\left| V_0(\epsilon )\right| ^{2}2z\langle sin^{2}k_{x}\rangle _{\epsilon }
\end{equation}

The contributions from the Coulomb-corrected vertex, on the other hand, require
the fourth-order term from Eq. (\ref{eq:VFour}): 
\begin{eqnarray}
\sum_{m}V_{m}^{(4)}(E_1,E_2,E_3,E_4) = 
\frac{1}{L^{3}}\sum _{\vec{k}_1,\vec{k}_2,\vec{k}_3,\vec{k}_4} \nonumber \\
\delta (\epsilon _{\vec{k}_1}-E_1)\delta (\epsilon _{\vec{k}_2}-E_2)\delta (\epsilon _{\vec{k}_3}-E_3)
\delta (\epsilon _{\vec{k}_4}-E_4) \times \nonumber \\
\sum _{m,m'}V_{m\sigma _1}(\vec{k}_1)V_{m\sigma _3}^{*}(\vec{k}_3)\sum _{m'}V_{m'\sigma _2}(\vec{k}_2)
V_{m'\sigma _4}^{*}(\vec{k}_4) \times \nonumber \\
\delta ^{*}(\vec{k}_1+\vec{k}_2-\vec{k}_3-\vec{k}_4)
\end{eqnarray}
 where \( \delta ^{*}(\vec{k}_1+\vec{k}_2-\vec{k}_3-\vec{k}_4) \) is
the Laue function, Eq. (\ref{eq:Laue_fnctn}). If we were to neglect momentum
conservation this expression would vanish identically.

Inserting the explicit expressions for the hybridization matrix elements Eq.
(\ref{eq:hybasa}) reduces to a sum of local contributions 
\begin{eqnarray}
\sum_{m}V_{m}^{(4)}(E_1,E_2,E_3,E_4)= 
\sum _{\vec{R}_n}\sum _{m,m'}I_{m\sigma _1}(\vec{R}_n,E_1) \times \nonumber \\
I_{m'\sigma _2}(\vec{R}_n,E_2)I_{m\sigma _3}^{*}(\vec{R}_n,E_3)
I_{m'\sigma _4}^{*}(\vec{R}_n,E_4)
\end{eqnarray}
 where the averages 
\begin{equation}
I_{m\sigma }(\vec{R}_n,E)=N(E)\langle V_{m\sigma }(\vec{k})e^{i\vec{k}\cdot \vec{R}_n}\rangle _{E}
\end{equation}
 are given by 
\begin{eqnarray}
I_{m\sigma }(\vec{R}_n,E)=N(E)V_0(E)\left( -i\right) \sqrt{4\pi }\sum _{\vec{R}_{j}}Y_{3m}^{*}\left( \hat{R}_{j}\right)
\nonumber \\ 
\langle e^{i\vec{k}\cdot \vec{R}_{j}}e^{i\vec{k}\cdot \vec{R}_n}\rangle _{E}\quad .
\end{eqnarray}
In the next step, we sum over the magnetic quantum numbers \( m \) and \( m' \)
\begin{eqnarray}
\lefteqn {\sum _{m=-\ell }^{\ell }I_{m\sigma _1}(\vec{R}_n,E_1)I_{m\sigma _3}^{*}(\vec{R}_n,E_3)} &  & \nonumber \\
 & = & V_0(E_1)V_0(E_3)\sum _{\vec{R}_{j},\vec{R}_{j}'}\langle e^{i\vec{k}_1\cdot 
\left( \vec{R}_{j}+\vec{R}_n\right) }\rangle _{E_1}\langle e^{-i\vec{k}_3\cdot 
\left( \vec{R}_{j'}+\vec{R}_n\right) }\rangle _{E_3} \nonumber \\
 & \times & 4\pi \sum _{m=-\ell }^{\ell }Y_{3m}^{*}\left( \hat{\vec{R}}_{j}\right) Y_{3m}^{*}\left( \hat{\vec{R}}_{j'}\right)
\nonumber \\ 
 & = & NV_0(E_1)V_0(E_3)\sum _{\vec{R}_{j}\neq 0}\langle e^{i\vec{k}_1\cdot \left( \vec{R}_{j}+
\vec{R}_n\right) }\rangle _{E_1} \nonumber \\
 & \times& \langle \left( e^{-i\vec{k}_3\cdot \vec{R}_{j})}-
e^{i\vec{k}_3\cdot \vec{R}_{j}}\right) e^{-i\vec{k}_3\cdot \vec{R}_n)}\rangle _{E_3}
\end{eqnarray}
 The local term \( \vec{R}_{j} \) vanishes identically due to the symmetry
of the constant energy surface. For finite \( \vec{R}_n \), the averages
over the constant energy surfaces decay with increasing \( \vec{R}_n \) due
to the oscillatory behavior of the integrand. The leading contribution to the
lattice sum involves the nearest neighbor term, e. g. \( \vec{R}_n=(1,0,0) \).
In the subsequent summation over the nearest neighbors \( \vec{R}_{j} \), the
nonvanishing contributions are given by \( \vec{R}_{j}=(-1,0,0) \) and \( \vec{R}_{j}=(1,0,0) \)
yielding 
\begin{equation}
\langle 1-\cos 2k_{x}\rangle _{E_1}\langle 1-\cos 2k_{x}\rangle _{E_2}=4\langle \sin ^{2}k_{x}\rangle _{E_1}\langle \sin ^{2}k_{x}\rangle _{E_2}\quad .
\end{equation}
 The leading contribution to the fourth-order term \( V^{(4)} \) hence factorizes
according to 
\begin{eqnarray}
\frac{1}{N}\sum_{m}V^{(4)}(E_1,E_2,E_3,E_4)= \nonumber \\
zN\prod _{r=1}^{4}N(E_{r})V_0(E_{r})2\langle \sin ^{2}k_{x}\rangle _{E_{r}}
\end{eqnarray}
 The average \( 2\langle \sin ^{2}k_{x}\rangle _{E} \) is of order unity as
mentioned above. If we neglect the dependence of the density of states and \( V_0 \)
on the energy \( E \) we find for the averaged value of \( V^{(4)}(E_1,E_2,E_3,E_4) \)
\begin{equation}
\label{eq:averagedV4}
V^{(4)}=\frac{N}{z}\left( \frac{N(\epsilon _f)\Delta}{\pi} \right) ^{2}
\end{equation}

In conclusion, we showed that in the leading approximation of the hybridisation
matrix element expansion, Eq. (\ref{eq:hybasa}), 
the ratio \( \frac{V^{(4)}}{(\Delta N(\epsilon _f)/\pi )^{2}}\simeq \frac{N}{z} \) i.e. of the order of unity.

\end{multicols}


\begin{thebibliography}{10}
\bibitem{CommentHybridizationWidth}
The non-degenerate case may evolve from the orbital degenerate case as the
  consequence of the crystal electric field splitting.
\bibitem{CommentOrbitalDegeneracy}
Both the degenerate ( \( N>2 \) ) and the non-degenerate cases are considered.
  In the case of the non-degenerate Anderson impurity index \( m \) consides
  with the electron spin index \( \sigma \) and the coupling \( V(\vec{k}) \)
  does not depend ( in a paramagnetic host and by neglecting the spin-orbit
  interaction ) on the spin orientation. So in the non-degenerate Anderson
  impurity case the subscript m coincides with \( \sigma \) and \( V_{m\sigma
  }(\vec{k})\rightarrow V(\vec{k}) \).
\bibitem{Commentnca}
As it is known from the NCA-method, Eq. (\ref{eq:1stiteration}) is the first step 
in the NCA iteration procedure ( see \cite{Bickers87b} ) which already captures right 
the exponent of the Kondo energetic scale.
\bibitem{CommentSigmah}
One has have in mind that Eq. (\ref{eq:efh}) is strictly speaking applyed for the bubble 
of Figure \ref{fig:BubbleSelf}, a. For Figure \ref{fig:BubbleSelf}, c $\sigma$ changes sign
`across' interaction $U$.
\end{thebibliography}
\end{document}